# THE E-INTELLIGENCE SYSTEM

Vibhor Gautam (vibhorgautam907@gmail.com)
Vikalp Shishodia (vikalpshishodia37@gmail.com)

## ABSTRACT

Electronic Intelligence (ELINT), often known as E-Intelligence, is intelligence obtained through the use of electronic sensors. Other than personal communications, ELINT intelligence is normally obtained. The goal is usually to determine a target's capabilities, such as the placement of radar. Active or passive sensors can be employed to collect data. A provided signal is analyzed and contrasted to collected data for recognized signal types. The information may be stored if the signal type is detected; if no match is found, it can be classed as new. ELINT collects and categorizes data. In a military setting (and others that have adopted the usage, such as business), intelligence is information that helps an organization make decisions that can provide them a strategic advantage over the competition. The term "intel" is frequently shortened. The two main subfields of signals intelligence (SIGINT) are ELINT and Communications Intelligence (COMINT). The terminologies are specified by the US Department of Defense, and the categories of data reviewed are used by intelligence communities throughout the world [1].

## I. INTRODUCTION

Electronic intelligence (ELINT) is information derived solely from electrical impulses and does not include speech or writing (which are considered COMINT). There are several primary branches to it.

Technical ELINT (TechELINT) [2] explains the structure of the signal, its emission characteristics, its methods of operating, and the functions of the emitter and weapon system affiliations of emitters such as guidance messages and detection systems. Perhaps one TechELINT main goal is to collect signal characteristics that may be used to figure out an emitter's capabilities and as part of a bigger system, such as a land radar identifying airplane, improved radio, counteraction, or refute gear has been developed. The phrase "electronic warfare" encompasses the whole process, as well as countermeasures.

Another key discipline is operating ELINT (OpELINT), which focuses on discovering particular ELINT targets and analysing system functioning trends. The "Electronic Order of Battle" is the name given to these outcomes (EOB). Another service given by OpELINT is threat evaluations, sometimes known as "tactical ELINT." On the battlefield, OpELINT intelligence tools aid military missions' planners and local army generals. The gathering, analysis, and dissemination of foreign telemetry signals intelligence was a prior third key branch of ELINT (TELINT).

Surveilling, analysing, and evaluating alien telemetry yields technological and espionage information, known as TELINT. Telemetry Intelligence was initially considered to be part of ELINT since TELINT (later termed FISINT — Foreign Instrumentation Signals Intelligence) operations are intimately tied to TechELINT processes. When foreign rockets 2 and space probes are being built and tested, TELINT is a crucial part of performance data. TELINT is capable of providing a wealth of operational data about international satellites and spaceships [2].

The Radar Systems are the centres of Electronic Intelligence's attention, as well as the analysis of their properties. Observation, gathering, advance detection, TWS, guidance systems tracking, GCI, and other types of combat system radar systems are all threats.

Signal types include pulses, pulse Delta, CW, ICW, modern radars employing PC (pulse & frequency coding), frequency and PRI nimble radar systems, AM, FM, and others.

Capability of the system
- Computer-controlled and software-programmable system.
- Accurate signal capture and data processing in a short amount of time.
- Accurate signal analysis and presentation of emitter activity based on the operator's selection of criteria, Fingerprinting and emission recognition (pulse analysis- intra and inter).
- Data library linkage and management, efficient radar location, and the ability to upload recorded data / processes data in near real time over a separate encrypted aerial data link.
- GPS time stamping capability for all recorded data every 1 second.

When used tactically, the Electronic Support Measure (ESM) package of a weapon system might include ELINT equipment, which allows for detection and passive geo-location.

For a more in-depth signal study activity, a technical analysis is necessary (capture and analysis of material, as well as identification off-line).

- **Tactical Operations:** The Military, Marine, and Air Wing all want to know where all opposing radars are and what they can do with them. The Radar Order of Battle is what it's called (ROB) [3].

  For example, in order to determine how to neutralize enemy radars or take suitable counters over them, the Air Force must first understand how and where opponent radars are built.

  Figure 1 is an example of tactical presentation.

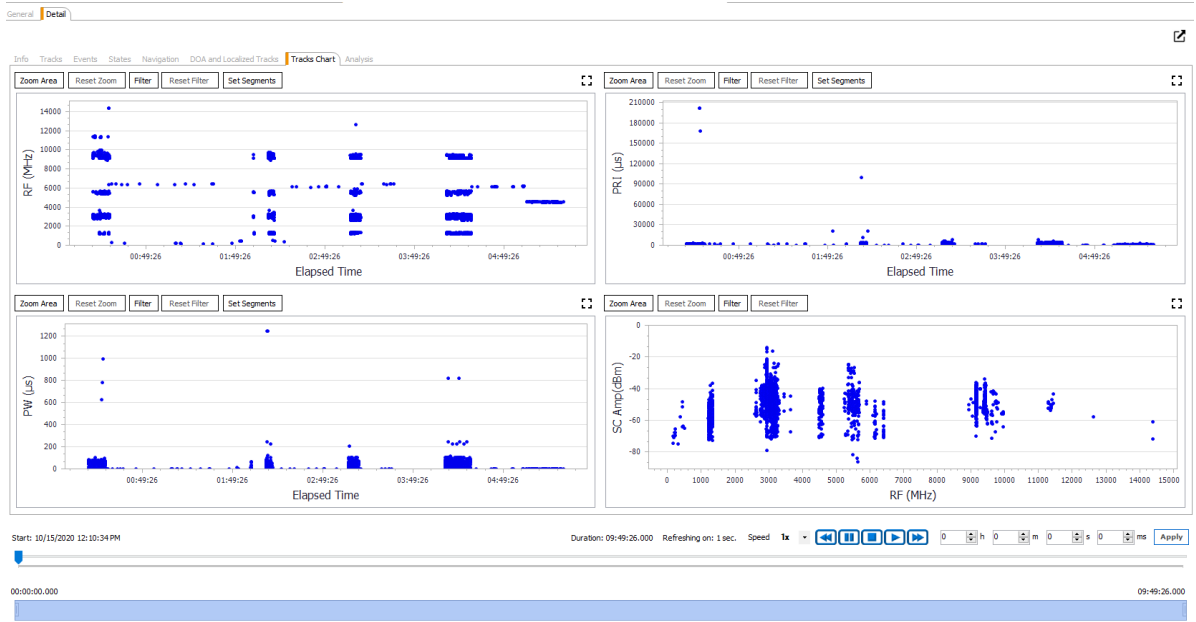

Figure 1: Theoretical Presentation of Tactical Data (ROB), Reproduced from [3]

- **Technological Operations:** Instead, intel attempts to acquire all radar operating modes and spectra from adversary radars. Understanding how defensive systems function, detecting new radars and operating modes, and keeping track of new adversary electronic advances and equipment are all required [3].

    Figure 2 is an example of a technological presentation.

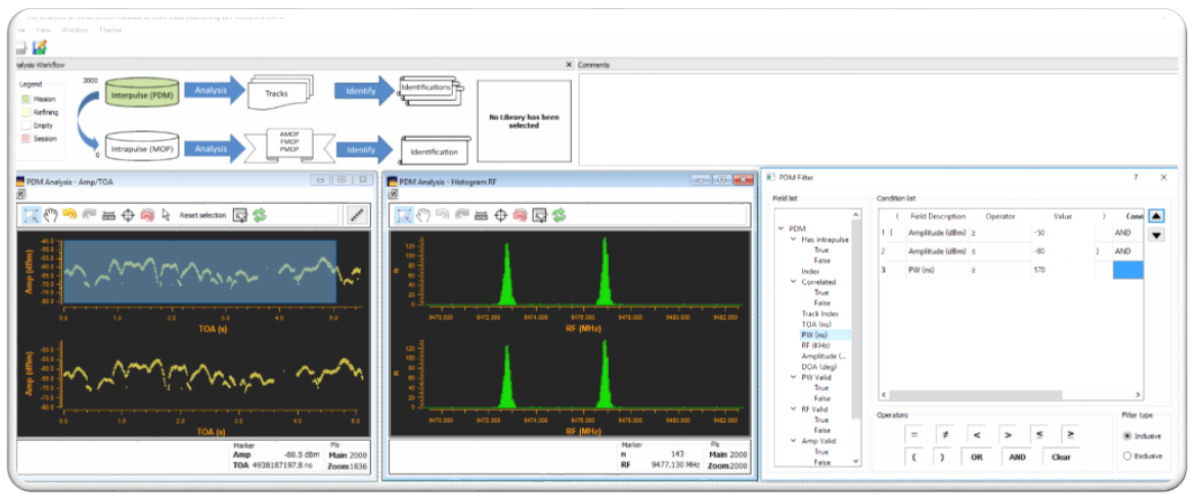

Figure 2: Theoretical Presentation of Technical Data
Reproduced from [3]

To back up the following many points of view
- Locate and identify hostile radar in a defined area of interest (AOI)
- Keep an eye on the whole electrical spectrum envelope
- Keep a safe distance from enemy radar installations;
- To achieve these operating criteria, in most ELINT systems, there are two kinds of receivers:
- A Receiver with a Panoramic View (Wide Band Receiver, WBR is employed to keep an eye on the surroundings, and
- A Selective Receiver is used to measure radar waveforms (Super Heterodyne Receiver, SHR).

The result of the ELINT Technical Analysis is the formation of combat systems database systems for radars affiliated with armaments (lookup and monitoring radars) and rocket claimants, that are then retrieved into the task datasets of the EW self-protection systems of warfare systems to aid in the detection phase.

## II. HISTORY

ELINT was born when the Allies and the Axis discovered and used radar during World War II. The British pioneered the earliest Allied ELINT activities, as recounted in Dr. R. V. Jones' book The Secret War.

Since most German interceptors were used to attack Allies planes over Germany at the time, the US Army Fighter Aircrafts were extremely keen in ELINT. They wanted to learn all they could about radars, such as how to avoid, "block," or "fake" them. During the second World War, ELINT was employed efficiently by all US army divisions against German land transponders, as well as Japanese aerial, navy, and underwater interceptors [2].

In early 1943, a B-24 bomber flew an ELINT operation above Kiska Peninsula inside the Aleutians to survey the Japanese radars upon that island. ELINT devices were placed aboard several B-24 bombers in 1944, and they have been used to spearhead the blocking of Nazi base radar systems across Europe.

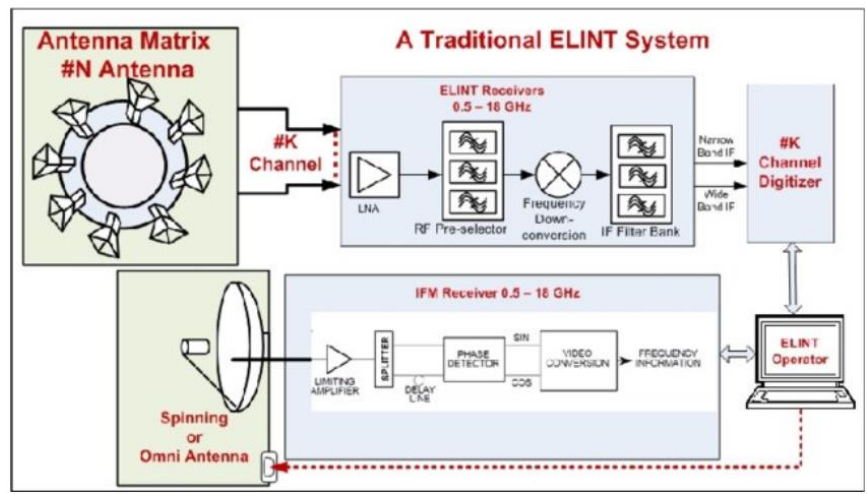

Figure 3: A traditional ELINT system
Reproduced from [4]

Soon after WWII, the United States Air Forces of European Union (USAFE) began an aggressive TechELINT and OpELINT programme, that included establishment of a robust and broad collaboration program with a number of NATO allies. The Army-Navy Electronics Evaluation Group (ANEEG) was established by the US Department of Defense in 1952, with a staff of roughly thirty individuals [2]. It was held at the Navy's Nebraska Avenue Naval Security Station (NSS). Despite its limited collecting capabilities, the ANEEG served as just a central analytical center for ELINT interception processing and interpretation, and a central figure for collaboration of tough ELINT analytic challenges.

## III. CURRENT STATE

### ➢ Role in air warfare

Radar intercept and knowledge of its locations and operating procedures is a common kind of ELINT. Attacking troops could be able to prevent radar coverage, or electronic warfare units may be able to disrupt radar systems or broadcast false signals if they are aware of their characteristics. A "soft kill" is when a military unit confuses a radar electronically, whereas a "hard kill" is when a military unit sends specialized missiles or bombs at radars. Radar homing guidance systems are included on several modern air-to-air missiles, which are designed to counter massive radars in the air [5].

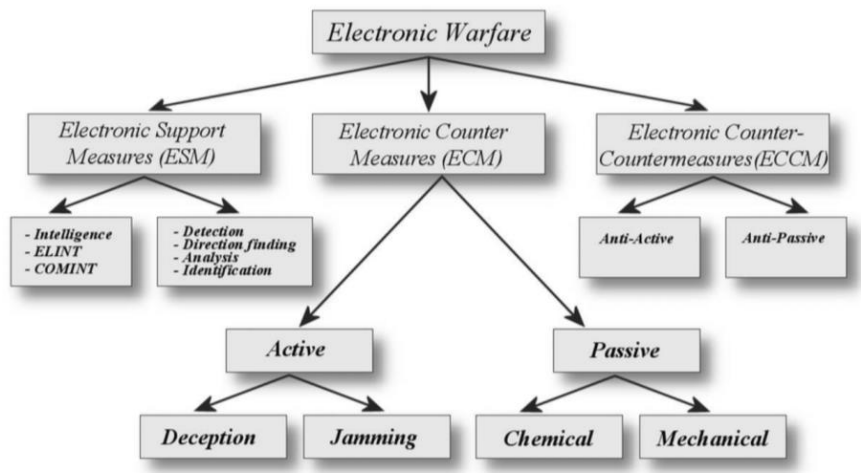

Figure 4: Elements of Electronic Warfare
Reproduced from [6]

Air strikes may be designed to stay away from the most strongly fortified locations and wing on an avionics suite that will offer the airplane the best chance of avoiding incoming fire and jet sweeps if they know in which each ground missiles and generally pro cannon systems is stationed and what kind it is. It may also be used to disrupt or spoof an adversary's defence network (also electronic warfare). Electronic intelligence is critical for stealth operations; stealth aircraft cannot operate in complete darkness and must understand which places to avoid. Conventional aircraft, therefore, must be aware of their location to disable or dodge stationary or quasi missile defence systems.

## ➢ UAV-Based ELINT

The resurgence of more traditional threat scenarios, along with the downing of a Malaysia Airlines planes, using a BUK technology, have started to draw media awareness to what has long been a recurring theme in our community: The danger of radar-guided surface-to-air missiles (SAMs). The ability of the SAM threat to create Pro, Areas Rejection (A2AD) domes in distressing areas throughout the globe have put the hard Suppression of Enemy Air Defense (SEAD) aim and methods into sharp focus [7].

When preparing to suppress the systems that make up a complex anti air (IAD) system inside the A2AD sphere, we quickly realize that a thorough ISR operation will be required to create and maintain the digital command structure (EOB) and temperament of the elevated detectors and affiliated military hardware.

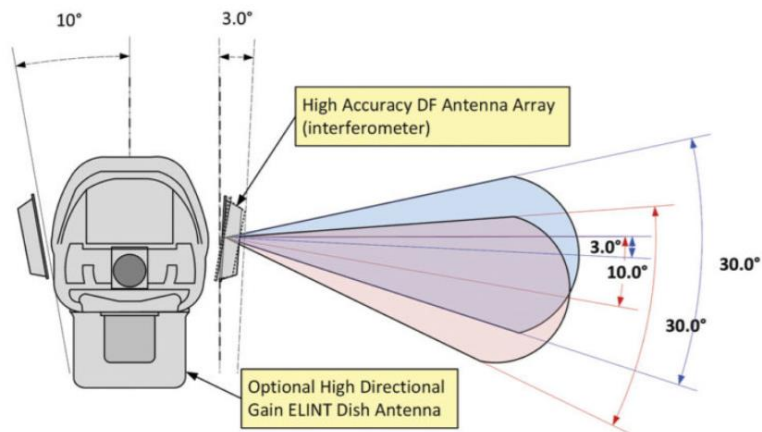

Figure 5: Antenna array installation considerations
Reproduced from [7]

The primary ISR task will be to efficiently deploy signals intelligence (SIGINT) assets capable of acquiring both electronic and communications intelligence (COMINT) on IAD network elements. Historically, manned platforms were the domain of ELINT, particularly strategic ELINT, and this is still the case today. Unmanned Aerial Systems (UAS) are increasingly being used for SIGINT missions (UASs).

*ELINT SYSTEM ARCHITECTURE APPROACH*

The structure and configuration of an ELINT-capable UAS as well as its aerial framework, also known as an unmanned aerial vehicle (UAV), must always be established through a careful process that includes knowledge and feedback from subject matter experts (SMEs) and technicians to months of actual operations and maintenance ELINT system design skills. The UAS ELINT architecture process includes establishing needs and urgencies, defining the Concept of Operations (CONOPS), evaluating the risk, and a range of operation characteristics and situations. [7].

## ➢ Tactical Exploitation System-Forward (TESFWD)

The intelligence (ELINT) capabilities of the TES-electronic FWD are well-known. For commanders, ELINT is essential because it offers intelligence on adversary radar and Electronics, Rocket, and Land Order of Battle. ELINT continues to give tactical information and indicators and alerts about danger nations' radar, ballistic, and land assault positions and intents. ELINT has assisted key wars in the past, including the Soviets and Operation Desert Shield, and will do so in the future. The risk of a high-intensity war still exists. In the future, ELINT will remain a few of the principal information specialties utilised for supporting significant military operations, just as it was during Operation Iraqi Freedom's planning and execution phase (OIF).

Some radars are also shifting to higher frequencies, far into the millimetre-wave range, posing new challenges. Other radars operate in a heavily packed EM region around RF and around 6 GHz, making spotting changes much more challenging and indicating a pattern that is likely to worsen as cellular telephone services increase their range and, in some cases, overlap numerous services along the same wavelengths. [8].

ELINT receivers leverage consumer-market technologies to address all of these issues. These include host processors with high sampling frequencies, precision, and dubious dynamic range, as well as ADCs and DACs have fast sampling frequency, precision, and a wide dynamic range with no spurious signals (SFDR).

## ➢ ELINT and ESM

ESM is an ELINT technique that make use of a variety of electronic surveillance equipment, however the phrase is only used in tactical combat. Following WWII, the RDF was extended to include ELINT using radar bandwidths and frequencies lower communications networks, ending in the NATO ESM family, which comprises ships equipped with US AN/WLR-1—AN/WLR-6 devices or aerial units. Electronic countermeasure is another name for EA (ECM). ESM offers data for electronic counter-countermeasures (ECCM), including detecting and modifying radar settings to prevent spoofing or jamming modes.

## ➢ Sampling of Radar ESM and ELINT Receivers

ELINT detectors aren't required to create split-second, survival decisions like radar warning receivers (RWRs), but it doesn't make their position any easier or one's layout somehow less complex. Since our last assessment two years ago, they've improved noticeably in terms of overall efficiency and their ability to stay up with the radars they are competing versus.

The tasks that an ELINT receiver must perform have never been easy, but they have become significantly more challenging in recent years. The service life of a radar system is generally calculated in decades instead of years. As a result, an ESM or ELINT system should encompass an ever increasing roster of radars, both older and newer, whose profiles might change with passing time as upgrades, parts are changed, and their efficiency alters as emerging innovations are introduced. [9].

Various types of radars possess different electro - magnetic identities, therefore a missile defense systems radar would be different from one another, such as rocket seekers, or even radars with same type might very well change through one system toward the next. Furthermore, today's duplexer, multifunctional radars integrate complex signal modulation methods with harmonic mobility, intra pulse waveforms variations, odd waveform such as pseudo sound modulate, very quick side-lobe changes, digitized beamforming, and a slew of other deceptive tactics.

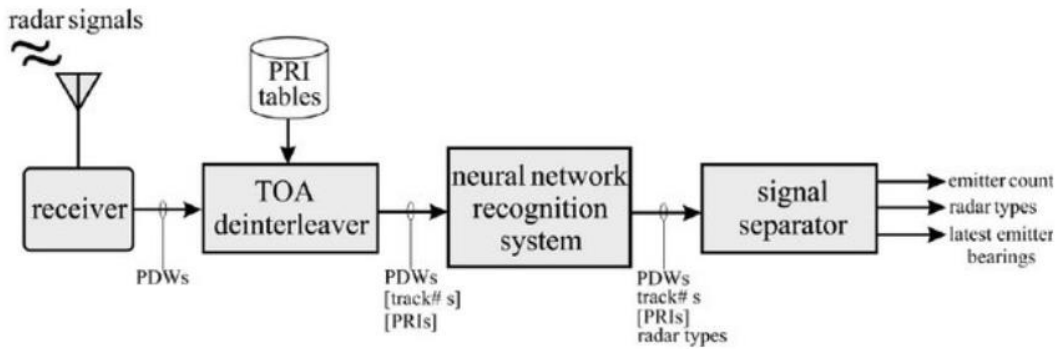

Figure 6: High level block diagram of a radar ESM
Reproduced from [10]

With earlier ESM and ELINT sensors, low-probability-of-intercept radars systems, like those that employ "frequency modulated continuous-wave" and "frequency-modulated interrupted continuous-wave", are significantly more difficult to detect, much less identify and assess. So summary is that, despite detecting a radar is feasible, explaining it might be challenging at times.

➢ **ELINT for Meaconing**

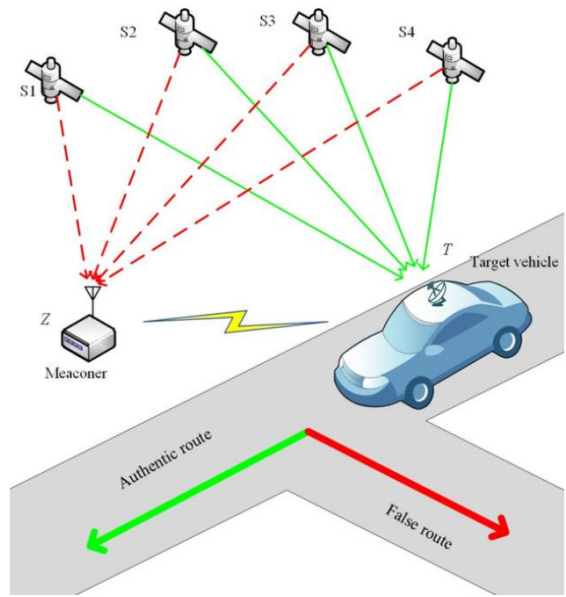

Figure 7: Basic principle of GNSS meaconing
Reproduced from [11]

Analysing the characteristics of opponent satellite systems, such as broadcasting beacon, then retransmitting misleading information using a combination of electronic warfare and intelligence is known as meaconing.

> **Foreign Instrumentation Signals Intelligence**

FISINT is a subcategory of SIGINT that focuses on inhuman communications. Telemetry (TELINT), video data linkages, and tracking systems are examples of foreign instrumentation signals. TELINT is a key component of a country's technical verification system for weapons control.

## IV. CONCLUSIONS AND FUTURE SCOPE

A current research effort is WolfPack, a ground-based cyber warfare system. A "pack" of "wolf" makes up WolfPack. Wolves are electronics monitoring nodes with the ability to locate and classify targets, and they can employ both ELINT and radiofrequency MASINT techniques.

The wolves might be delivered by hand, artillery, or airdrop. WolfPack, as well as Distributed Suppression of Enemy Air Defenses (DSEAD), a development of SEAD, might be used in a new counter-ESM sub-discipline developed by the Air Force. If the Wolves are near to the target and have jammers or other ECM, they won't require much power to mask the signatures of allied ground soldiers on frequencies critical for communication or local detection. DSEAD is comparable to DSEAD, notwithstanding the fact that it operates at radar frequencies. It could be interesting to compare and contrast this counter-ELINT activity with electronic countermeasures [12].

The bulk of countries are spending inorganically since their SIGINT systems have not yet been built. The integration of COMINT and ELINT using a variety of platforms will be the main priority in the future years, with a focus on building policies and procedures to put the whole SIGINT system under one roof.

In the United States and Europe, ELINT is becoming more important, and combining COMINT with ELINT is a growing field of technical and operational strategy. With escalating tensions in the South China Sea, the US is stepping up its efforts to develop maritime SIGINT. The US and NATO countries will witness a reduction in ground-based tactical SIGINT requirements once they return from the Afghanistan conflicts. In the future, electronic countermeasures such as space warfare and unmanned aerial aircraft will be utilized to combat wars.

For SIGINT and electronic warfare, China is substantially investing in spacecraft and unmanned aerial vehicles (UUVs). Other nations have suffered enough as a result of China's actions in the South China Sea. Japan and India are establishing their own SIGINT strategy to counter China's aggressive efforts, which will include additional satellite projects, SIGINT aircraft, and marine SIGINT. As a result, demand for SIGINT in Asian markets will increase.

By 2029, the SIGINT industry will have grown to $23.42 billion, increased from $15.75 billion in 2021. Between 2023 and 2027, ground-based SIGINT, unmanned and space aerial systems, and marine SIGINT will account for the bulk of SIGINT deployments, resulting in considerable market growth.